# Unleashing 5G Seamless Integration with TSN for Industry 5.0: Frame Forwarding and QoS Treatment

OSCAR ADAMUZ-HINOJOSA[1,2], FELIX DELGADO-FERRO[1,2], JORGE NAVARRO-ORTIZ[1,2], PABLO MUÑOZ[1,2], PABLO AMEIGEIRAS[1,2]

[1]Department of Signal Theory, Telematics and Communications, University of Granada, Granada, Spain
[2]Research Center on Information and Communication Technologies, University of Granada, Spain

CORRESPONDING AUTHOR: OSCAR ADAMUZ-HINOJOSA (e-mail: oadamuz@ugr.es).

This work has been financially supported by the Ministry for Digital Transformation and of Civil Service of the Spanish Government through TSI-063000-2021-28 (6G-CHRONOS) project, and by the European Union through the Recovery, Transformation and Resilience Plan - NextGenerationEU. Additionally, this publication is part of grant PID2022-137329OB-C43 funded by MICIU/AEI/ 10.13039/501100011033 and part of grant FPU20/02621 funded by the Spanish Ministry of Universities.

**ABSTRACT** Integrating Time-Sensitive Networking (TSN) and 5th Generation (5G) systems is key for providing wireless low-latency services in industry. Despite research efforts, challenges remain. Due to the lack of commercial 5G modems supporting Ethernet-based sessions, tunneling mechanisms must be used to enable Layer 2 connectivity between TSN islands via IP-based 5G modems. Furthermore, harmonizing traffic classification and prioritization between TSN and 5G technologies is crucial for meeting industrial service requirements. In this work, we propose a Virtual Extensible LAN (VxLAN)-based solution to harmonize frame forwarding and Quality of Service (QoS) treatment among 5G and TSN. Our solution supports multiple Virtual Local Area Networks (VLANs) across several production lines. Furthermore, it supports TSN traffic mapping into 5G QoS flows. We use a 5G testbed to validate the effectiveness of the adopted solution. Our results show the average delay introduced by the proposed mechanisms is approximately 100 $\mu s$, which is significantly lower than the typical 5G packet transmission delay. Moreover, our findings demonstrate our solution preserves QoS treatment between the 5G system and TSN, ensuring that the priority of 5G QoS flows aligns with the priorities of industrial traffic flows.

**INDEX TERMS** 5G, industry 5.0, TSN, QoS, testbed, VxLAN.

## I. INTRODUCTION

The adoption of recent advances in IoT, Cyber-Physical Systems, cloud computing and Artificial Intelligence into industrial production is envisioned to revolutionize traditional industries, forming the foundation of Industry 4.0 [1]. Building upon this momentum, Industry 5.0 emerges as the next evolutionary step, seamlessly blending automation with human ingenuity to foster a more intelligent, sustainable, and personalized manufacturing ecosystem [2], [3]. This new paradigm enhances the hyperconnectivity established by Industry 4.0, elevating industrial operations by integrating robotic precision with human creativity through collaborative robots and advanced cyber-physical interactions.

Realizing the full potential of Industry 5.0 network platforms that provide efficient, low latency, reliable, and deterministic communications among all components. In this respect, the Institute of Electrical and Electronics Engineers (IEEE) introduced Time-Sensitive Networking (TSN), a set of standards enhancing industrial networks with synchronization, stream reservation, traffic shaping, scheduling, preemption, traffic classification, and seamless redundancy [4]. However, wired industrial networks face scalability and flexibility limitations due to the complexity of adding/relocating wired equipment, restricting mobility and coverage to cabled areas.

To overcome these issues, researchers explore the integration of the 5th Generation (5G) system as TSN bridges within TSN networks [5]. In this setup, production lines connect wirelessly to the 5G system, which then interfaces with the enterprise edge cloud through TSN bridges [6]. However, such integration poses the following challenges [4], [7]–[9]:

C1. Optimizing Ultra-Reliable and Low-Latency Communications (uRLLC) configuration in 5G networks is key for ensuring low-latency and high-reliability communication in industrial environments.
C2. Accurate synchronization signal transmission via the 5G radio interface. It is complicated due to the wireless channel variability and the User Equipment (UE) mobility.
C3. Maintaining deterministic packet scheduling in an integrated 5G-TSN network is difficult due to the latency and jitter introduced by the wireless transmission.
C4. TSN operates at L2 using Ethernet features like Virtual Local Area Network (VLAN) and Priority Code Point (PCP), while commercial 5G terminals only support IP-based sessions, even though 3rd Generation Partnership Project (3GPP) standards allow Ethernet-based sessions.
C5. TSN and 5G have different Quality of Service (QoS) architectures, making it challenging to harmonize QoS treatment for industrial traffic.

In our study, we focus solely on challenges C4 and C5. Specifically, it is crucial to enable Ethernet frame transmission between TSN islands via IP-based 5G systems while aligning QoS treatment among TSN and 5G. Under this context, some researchers [10]–[12] use Virtual Extensible LAN (VxLAN) to encapsulates Ethernet frames within IP datagrams, preserving Ethernet headers when transmitting TSN frames through 5G. However, these studies lack detailed guidelines on VxLAN configuration for correct Ethernet frame forwarding in scenarios where VLANs are shared across production lines. Furthermore they do not address QoS integration. Others authors [13], [14] propose mapping standardized 5G QoS Identifiers (5QIs) to TSN flows based on PCP values to harmonize QoS between 5G and TSN. However, they do not provide details on how to implement this mapping, especially considering that commercial 5G systems only support IP sessions. Furthermore, their approach overlooks VxLAN's structure; VxLAN encapsulation hides the Ethernet header from 5G's packet filters, making it difficult to map TSN flows to 5G QoS flows.

In this paper, we present an overview of frame forwarding and QoS treatment in an IP-based 5G system integrated with an industrial TSN network. We identify key requirements to harmonize these mechanisms, highlighting gaps in 5G-TSN interoperability. To address them, we propose a VxLAN-based solution which is validated through a testbed with a commercial 5G system. We demonstrate VxLAN encapsulation mechanisms introduce an average latency of approximately 100 $\mu s$, which is negligible compared to the packet transmission delays in commercial 5G systems (ranging from a few to tens of milliseconds) [15]. Moreover, our solution preserves QoS treatment between the 5G system and TSN, ensuring the priority of 5G QoS flows remains aligned with the PCP priorities of industrial traffic flows.

The paper is structured as follows. In Section II, we first provide an overview of TSN industrial networks and the 5G architecture, with focus on their integration. We then analyze the requirements for integrating 5G and TSN, specifically addressing frame forwarding and QoS treatment. In Section III, we present our proposed solution. Section IV describes the proof of concept and presents the performance results. Finally, Section V concludes the paper, summarizing the key findings and outlining potential areas for future work.

## II. INTEGRATING TSN INDUSTRIAL NETWORKS AND 5G: OVERVIEW AND REQUIREMENTS
### A. TSN INDUSTRIAL NETWORKS
As depicted in Fig. 1, there are three connectivity segments in a TSN-based industrial network [6]:

- *Edge Room*: serves as a centralized management segment, hosting control functions like Programmable Logic Controllers (PLCs) and cloud-based applications.
- *Production Lines*: include multiple field devices and potentially local PLCs for distributed control.

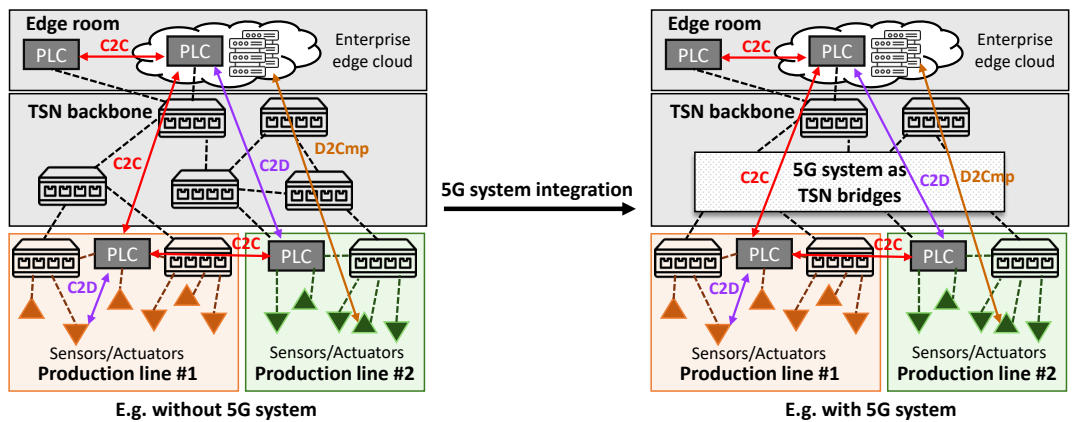

**FIGURE 1.** TSN-based industrial network in a single-site factory.

- *Industrial backbone*: connects the entire factory through TSN bridges and may also replace traditional communication technologies like PROFIBUS [16] within production lines.

**Communication Scenarios and Traffic Types.** In its effort to integrate a 5G system as TSN bridges (see Fig. 1) within TSN-based industrial networks, the 3GPP TS 22.104 (v19.2.0) [20] identified various types of communications: Controller-to-controller (C2C) communication between PLCs; Controller-to-device (C2D) communication between PLCs and field devices; and Device-to-computer (D2Cmp) communication between field devices and edge cloud servers. These scenarios involve traffic flows with distinct performance requirements as described in Table 1 [6], [17]:

- *Network Control*: manages tasks like time synchronization, network redundancy, and topology detection.
- *Cyclic Synchronous*: coordinates regular synchronized user plane data exchanges between devices.
- *Cyclic Asynchronous*: involves periodic but unsynchronized user plane data exchanges.
- *Events*: trigger messages based on metric changes.
- *Mobile Robots*: include movement control, task assignment, and sensor data.
- *Augmented reality*: provides real-time video and maintenance instructions.
- *Configuration and diagnostic*: handle non-critical data like device configuration and firmware downloads.

**Traffic Prioritization and Packet Scheduling.** Traffic prioritization is essential for managing frame transmission scheduling in TSN-based industrial networks. To that end, these networks rely on IEEE 802.1Q, which incorporates the PCP field (3-bit) in VLAN tags to prioritize frame transmission across layer 2 links. Different traffic types receive distinct PCP values based on their QoS requirements:

- *High-priority traffic (PCPs 4-7)*: critical frames, where data loss could precipitate critical malfunctions.
- *Medium-priority traffic (PCPs 2-3)*: critical frames, but if lost, data can still be recovered through frame retransmission.
- *Low-priority traffic (PCPs 0-1)*: non-critical frames.

These PCP tags, along with other metadata, classify packets into different queues which are prioritized by packet scheduling mechanisms (e.g., IEEE 802.1Qbv) at each egress port of every TSN bridge. The scheduling mechanisms ensure that frames in higher-priority queues are forwarded with precedence over those in lower-priority queues. Table I shows the PCP assignments proposed by 5G-ACIA [6].

The number of traffic flows with differentiated QoS is limited to the eight values defined by the PCP field. Therefore, although the number of individual traffic flows—each defined by a unique source and destination addresses—may exceed eight in industrial environments, QoS treatment across the entire TSN network is still restricted to these eight PCP values.

### B. 5G FEATURES FOR AN INTEGRATED 5G-TSN NETWORK

**5G-TSN Scenario.** We consider an industrial scenario with multiple production lines as depicted in Fig. 2. Each production line has a head-of-line TSN bridge providing L2 con-nectivity among field devices and a local PLC. This bridge connects to a single UE, enabling wireless access to other production lines, the edge room, or external networks via a 5G system. The edge room houses centralized PLCs and the edge cloud. Following 3GPP TS 23.501 (v19.0.0) [21], the 5G system integrates into the TSN network as virtual TSN bridges, with User Plane Functions (UPFs) and UEs as endpoints. Components such as Network-side Translator (NW-TT) and Device-side Translator (DS-TT) enable TSN translation functionalities. We focus on the data plane, excluding control plane functions for clarity.

TABLE 1. Industrial automation traffic types, performance requirement, and example of PCP mapping into 5G QoS flows

| Traffic Types [6] | Communication Scenarios [17] | Periodic / Sporadic | E2E Delay bound (ms) | Typical Data Size (Byte) | PCP | 5G QoS flows (3GPP TS 23.501 v18.5.0 - Table 5.7.4-1 [18]) | | |
|---|---|---|---|---|---|---|---|---|
| | | | | | | 5QI | Default Priority Level | Packet Delay Budget (ms) |
| Network Control | C2C and C2D | Periodic | [50, 1000] | Variable [50, 500] | #7 | #69 / #65 / #67 | #5 / #7 / #15 | 60 / 75 / 100 |
| Isochronous | C2D | Periodic | [0.1, 2] | Fixed [30, 100] | #6 | Non-standardized 5QI [13], [19]. | | |
| Cyclic Synchronous | C2C and C2D | Periodic | [0.5, 1] | Fixed [50, 1000] | #5 | Non-standardized 5QI [13], [19]. | | |
| Cyclic Asyncrhonous | C2C and C2D | Periodic | [2, 20] | Fixed [50, 1000] | #5 | #86 / #82 / #90 | #18 / #19 / #25 | 5 / 10 / 20 |
| Events | C2D and D2Cmp | Sporadic | [10, 2000] | Variable [100, 1500] | #4 | #87 / #88 / #89 / | #25 / #25 / #25 / | 5 / 10 / 15 |
| Mobile Robots | C2D and C2Cmp | Both | [1, 500] | Variable [64, 1500] | #3 | #90 / #3 / #71 | #25 / #30 / #56 | 20 / 50 / 150 |
| Augmented Reality | D2Cmp | Both | 10 | Variable [64, 1500] | #2 | #80 | #68 | 10 |
| Configuration and Diagnostic | C2C, C2D and D2Cmp | Sporadic | [10, 100] | Variable [500, 1500] | #1 | #7 | #70 | 100 |
| Best Effort | D2Cmp | Sporadic | N.A. | Variable [30, 1500] | #0 | #9 | #90 | 300 |

Although not depicted in Fig. 2, the considered scenario may also include UEs with mobility requirements, such as factory workers or mobile robots.

**5G QoS Flows.** 5G systems manage different data traffic types with specific performance requirements. 3GPP TS 23.501 (v19.0.0) [21] defines 5G QoS flows, each identified by a unique QoS Flow ID (QFI), facilitating traffic management based on quality requirements. The QFI is linked with descriptors such as 5QI (performance attributes) and Allocation and Retention Priority (ARP) (flow acceptance and preemption details). Each QoS flow belongs to a Packet Data Unit (PDU) session, which links a UE to a Data Network (DN). To map incoming traffic to a specific QoS flow, the 5G system uses a Packet Detection Rule (PDR) algorithm located in the UPF/UE. This algorithm analyzes IP/Ethernet header information based on a defined packet filter set, assigning each packet to the appropriate QFI.

**5QI Parameters.** 5QI is a value defining QoS features like resource type, priority level, packet delay, error limit, burst data volume, and averaging timeframe. These attributes dictate resource scheduling, transmission queue management, and protocol configurations across the 5G system. In the Radio Access Network (RAN), 5QIs guide traffic flow treatment via Data Radio Bearers (DRBs). One or more QoS flows are mapped onto one DRB, each tailored to meet specified QoS parameters throughout the 5G New Radio (NR) protocol stack.

Table 1 exemplifies a mapping of 5QIs to industrial traffic flows, aligning 5QIs with corresponding PCP values. We sort 5QIs based on their default priority levels, where lower numerical values indicate higher priority. These are then matched to PCP values, with higher-priority 5QIs paired with higher-priority PCP values. Only 5QIs with packet delay budgets suitable for each traffic type are included, ensuring compliance with both priority and delay requirements for industrial traffic.

Note that no 5QIs have been specified for isochronous and cyclic synchronous traffic types, as the 3GPP has not yet standardized any 5QI with a packet delay budget below 5 ms. This limitation is critical for isochronous and cyclic synchronous traffic in network scenarios with traffic congestion, where the 5G system could not guarantee packet transmission delays below 5 ms. Therefore, the 3GPP should consider standardizing new 5QIs in future standard releases to address the requirements of isochronous and cyclic synchronous traffic types.

### C. REQUIREMENTS FOR HARMONIZING FRAME FORWARDING AND QOS TREATMENT IN 5G-TSN NETWORKS

To the best of our knowledge, 5G market-available modems only supports IP-based sessions. This limitation poses a challenge for transmitting Ethernet frames within a 5G-TSN network, as the Ethernet header is removed when the frame enters the 5G system. To address it, several requirements must be met:

**Functional Requirements**:

R1. The 5G system must retain the Ethernet header of ingress TSN frames to support multiple VLANs and different traffic types classified by PCP values.
R2. The 5G system must support the transmission of Ethernet broadcast frames, essential for broadcasting across production lines or when the destination MAC address is unknown.

**Traffic Management and Flow Mapping Requirements**:

R3. Traffic classification and prioritization must be consistent between TSN and 5G. This requires mapping between TSN's PCP values and 5G QoS flows. It also involves the 5G system's packet filters must identify industrial traffic flows and map them to appropriate 5G QoS flows.

**Performance Requirements**:

R4. The 5G system must meet strict performance requirements for End-to-end (E2E) packet transmission latency [9], depending on the specific service.

### III. SEAMLESS FRAME FORWARDING AND QOS TREATMENT IN 5G-TSN NETWORKS
### A. ADOPTED VXLAN-BASED SOLUTION

Industrial traffic flows are typically identified by the tuple {VLAN ID, PCP} as Fig. 2 shows (see table with traffic flows generated by DN1 and DN2). Each production line may require multiple VLANs for isolating services (e.g., VLANs 100, 200 and 300 in production line #N), and VLANs can be reused across production lines (e.g., VLAN 100 and VLAN 200 in production lines #1 and #N). Additionally, industrial traffic flows may have different priorities distinguished by the PCP value, e.g., traffic flows with PCP 7 and PCP 5 within VLAN 100 (i.e., those represented in red).

As in [10]–[12], we adopt VxLAN to retain the Ethernet header of ingress TSN frames in the 5G system. VxLAN is a technology used to create a virtualized network overlay by encapsulating Ethernet frames inside IP packets. Essentially, VxLAN establishes an IP tunnel between two or more endpoints, known as VxLAN Tunnel End Points (VTEPs).

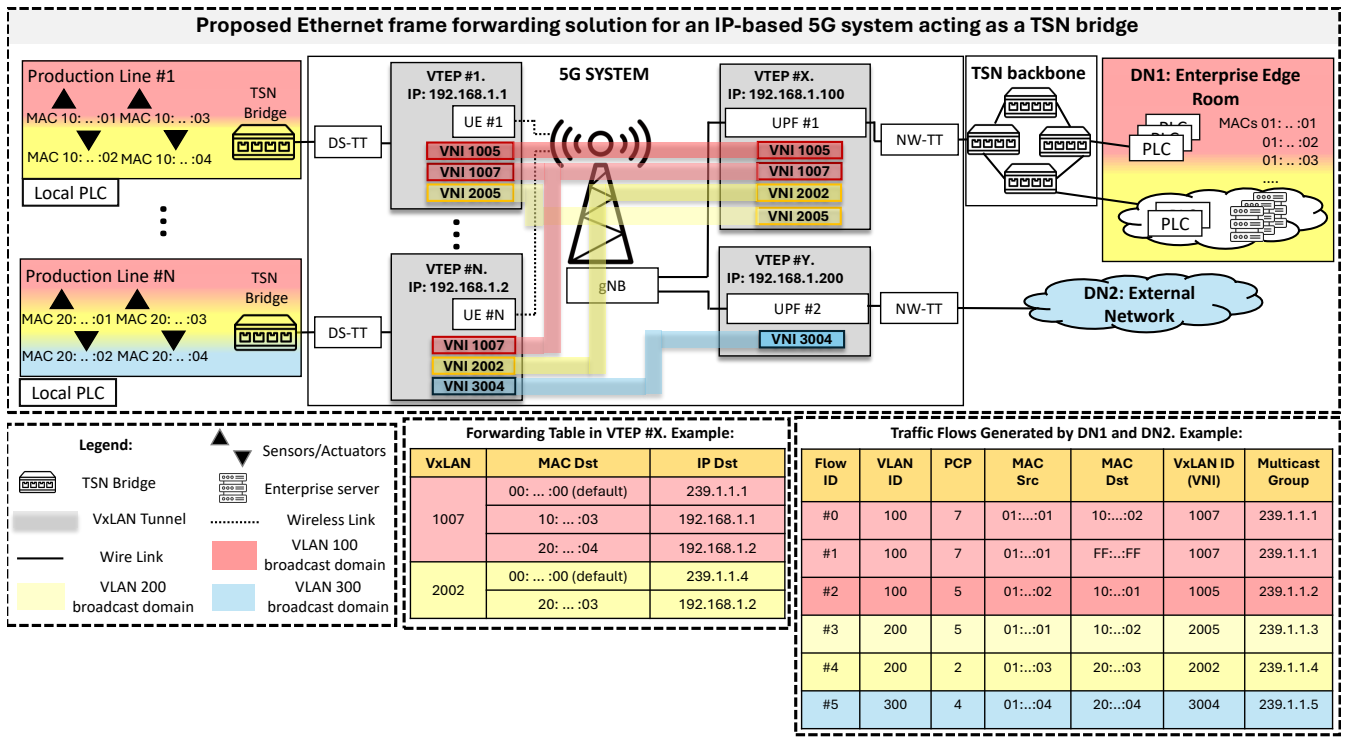

FIGURE 2. Seamless integration of the 5G system into the industrial network via VxLANs.

This tunnel allows Ethernet frames to be transported over an IP network –the 5G system– allowing Ethernet connectivity across separate TSN islands. Each VxLAN tunnel is identified by a Virtual Network Identifier (VNI), and each VTEP has a unique IP address. As shown in Fig. 2, for example, a VxLAN tunnel with VNI 3004 (blue tunnel) is established between VTEP #N (UE #N) and VTEP #Y (UPF #2).

**Advantages of VxLAN tunneling**. VxLAN encapsulation is essential to ensure Ethernet frame forwarding over IP-based 5G networks. Although, this approach comes with a trade-off: an additional overhead of 36 bytes per packet due to the external VxLAN, UDP, and IP headers, VxLAN tunneling provides several key benefits in 5G-TSN-based industrial network, including:

- *Interoperability and Multi-Vendor Integration*: VxLAN tunnels facilitate seamless operation and integration across diverse vendors, ensuring compatibility in multi-vendor environments. For example, some 5G devices and/or 5G network functions from a vendor may support Ethernet PDUs, while devices and network functions from other vendors may not.
- *Ease of Integration with Virtualization and Cloud Technologies*: VxLAN design aligns naturally with modern virtualization and cloud infrastructures, enabling dynamic and efficient network scaling.
- *Mature and Reliable Technology*: VxLAN has proven to be robust and less prone to failure, offering stability and reliability for network operations.
- *High Scalability*: VxLAN offers impressive scalability, with its 24-bit VNI field allowing the definition of up to 16777216 unique VxLAN tunnels. This feature makes it an ideal solution for networks with rapidly growing demands, as it can easily accommodate a large number of isolated broadcast domains.
- *Simplified Configuration and Monitoring*: The availability of existing tools for VxLAN ensures that configuration and monitoring processes are straightforward and user-friendly.
- *No UE Terminals Supporting Ethernet-based sessions*: To the best of the authors' knowledge, there are currently no UE terminals on the market that support Ethernet PDU sessions.

**Gaps for Frame Forwarding and QoS Treatment:** Despite VxLAN's benefits, this technology does not ensure seamless frame forwarding and QoS handling in a 5G-TSN network. One gap is VxLAN does not guarantee the 5G packet filter set can read the VLAN ID and PCP of an encapsulated Ethernet frame. These filters may treat the Ethernet frame as the payload of the VxLAN-encapsulated packet, blocking access to the Ethernet header fields. Additionally, the RFC 7348 (VxLAN standard) does not mandate that VLAN and PCP tags are preserved during VxLAN encapsulation, so this information may be lost. Other gap is VxLAN does not ensure broadcast/multicast frame transmission, which is crucial when a VLAN is shared across multiple production lines.

To address these gaps, we adopt a VxLAN solution which exhibits the following characteristics:

**VTEP Placement within the 5G System**: A VTEP is placed between each TSN translator, i.e, DS-TT or NW-TT, and each 5G endpoint, i.e., UE or UPF. They may also be integrated within the UEs (i.e, VTEPs #1 and #N) and the UPF (i.e, VTEPs #X and #Y) as Fig. 2 shows. The IP addresses of the VTEPs are those of the respective UE or UPF, e.g., 192.168.1.2 is the IP address of UE #N and, thus, VTEP #N.

**Mapping Industrial Traffic into VxLANs**: When a frame enters the 5G system through an ingress VTEP (i.e., UE or UPF), it is encapsulated in a VxLAN using a unique VNI based on the tuple {VLAN ID, PCP}. Therefore, establishing a direct correspondence between each VNI and the tuple {VLAN ID, PCP} is essential for restoring these tags during VxLAN decapsulation and enabling their utilization in the TSN network beyond the egress VTEP. For example, when a frame with the tuple {VLAN 200, PCP 2} enters UPF #1 (i.e., VTEP #X), it is mapped into VNI 2002, which is formed by concatenating the VLAN ID digits followed by the PCP digit.

The number of VLANs in an industrial environment depends on the required independent broadcast domains, limited to 4096 due to the 12-bit VLAN ID field. Within each VLAN, up to eight industrial traffic flows can be differentiated in terms of QoS treatment, with each flow distinguished from the others by the 3-bit PCP field. This gives a total of 32768 possible VLAN and PCP combinations. Our solution maps each combination to a unique VxLAN tunnel, requiring at most 32768 tunnels. This is fully supported by the 24-bit VNI, which allows for over 16.7 million unique tunnels, ensuring scalability regardless of the number of industrial traffic flows and TSN islands in the industrial environment.

**Frame Forwarding if MAC destination is known**: When a frame (e.g., from flow #4) enters the 5G system (e.g., through UPF #1), if the ingress VTEP (e.g., VTEP #X) knows how to reach its destination MAC address, this VTEP looks up its forwarding table (e.g., see mapping of MAC 20:...:03 into VxLAN 2002) to determine the IP address of the egress VTEP (e.g., VTEP #N's IP address is 192.168.1.2). It then forwards the encapsulated frame to the egress VTEP (e.g., VTEP #N) to later reach the destination MAC (e.g., actuator from Production Line #N).

**Frame Forwarding if MAC destination is unknown**: When the ingress VTEP (e.g., VTEP #X) receives an incoming unicast frame (e.g., VLAN 100 and PCP 7 from traffic flow #0) for which it does not know how to reach its destination MAC address (e.g., 10:...:02, which does not appear in the VTEP #X's forwarding table entry for VxLAN 1007), the frame must be broadcast across the VLAN broadcast domain (e.g., production lines #1 and #N for VLAN 100). Thus, when the frame enters the 5G system, several egress VTEPs (one per production line) could be a possible destination. To transmit such frame across the VLAN broadcast domain, we associate an IP multicast group[1] with the VxLAN mapping the corresponding {VLAN ID, PCP} tuple (e.g., 239.1.1.1 group with VNI 1007 as defined in the Fig. 2's traffic flow table). Therefore, for transmitting this frame, the ingress VTEP uses the default entry in its forwarding table (i.e., an all-zeros MAC address in the VTEP #X's forwarding table). That means the encapsulated frame is forwarded to all the VTEPs belonging to the same multicast group (e.g., VTEP #1 and VTEP #N). Note that *backward learning* is used in VTEPs to learn how to reach a destination MAC address, thereby associating that MAC address with the IP of an egress VTEP. The same principles apply when a broadcast frame (e.g., destination MAC FF:...:FF as seen in frames from traffic flow #1) enters the ingress VTEP (e.g., VTEP #X).

**Mapping TSN Flows into 5G QoS Flows**: To ensure consistent traffic classification and QoS treatment in both 5G and TSN, ingress IP packets in the 5G system are mapped into specific QoS flows using the packet filter set, which is located at the UE (for uplink traffic) or the UPF (for downlink traffic). We consider the use of the Differentiated Services Code Point (DSCP) field in these filters, proposing the ingress VTEP sets the outer IP header's DSCP field based on the incoming traffic's PCP value. Each VxLAN is linked to a specific PCP value, with the ingress VTEP checking the VNI to set the DSCP value accordingly. Fig. 3 includes a table for mapping PCP values to DSCP values [23].

To achieve the highest possible granularity in terms of QoS differentiation, our approach ensures a one-to-one mapping between the QoS differentiation levels in the industrial network—determined by PCP values—and 5G QoS flows. In TSN, traffic differentiation is based on PCP values, which define up to eight distinct QoS levels (0 to 7). To achieve this granularity, each of these QoS levels is mapped to a separate 5G QoS flow. This is feasible within the 5G system, as it defines significantly more than eight different 5QI values (see 3GPP TS 23.501 v18.5.0 - Table 5.7.4-1 [18]), ensuring that the level of granularity in the 5G system is not a limiting factor compared to the TSN network. To carry out such mapping, each 5G packet filter must associate a specific DSCP value with an unique QFI, which describes a specific 5G QoS flow.

However, it is also possible to map multiple industrial traffic flows with different PCP values into a single 5G QoS flow. This approach, while feasible, results in a loss of QoS differentiation in the 5G system, as multiple industrial traffic flows with distinct performance requirements would be aggregated into the same 5G QoS flow. This mapping can be implemented by configuring multiple 5G packet filters that associate different DSCP values (derived from different PCP values) with the same QFI. This mechanism allows multiple industrial traffic flows to share a common

---

[1]5G systems support Multicast–Broadcast Services (MBS), introduced in 3GPP TS 23.247 (v18.6.0) [22].

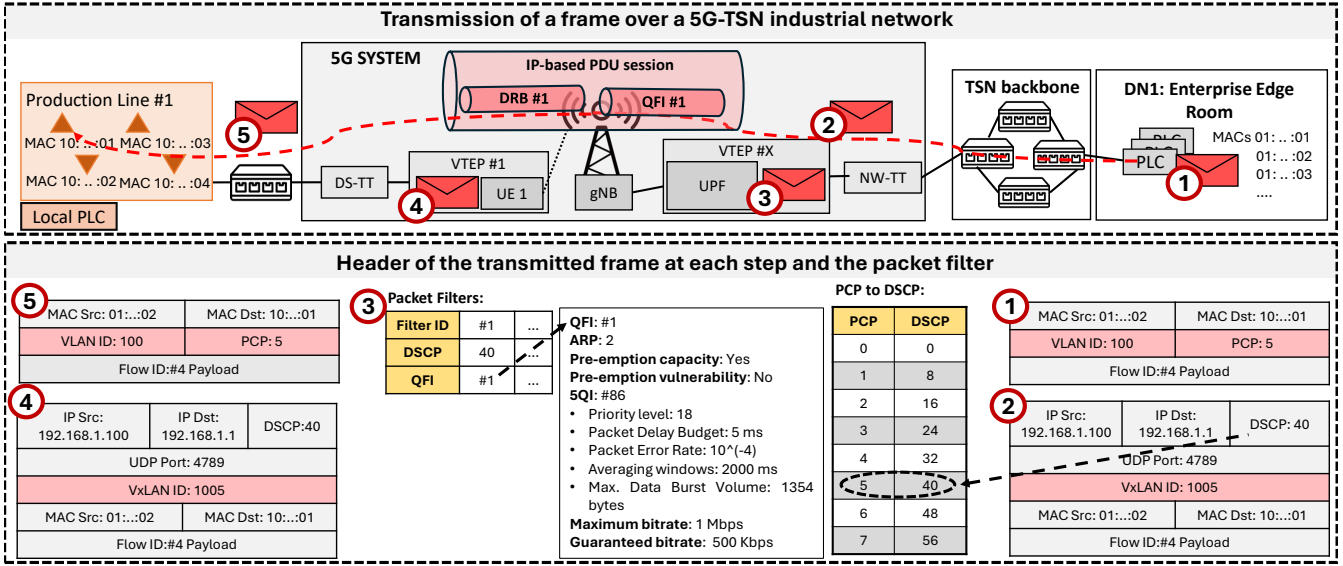

**FIGURE 3.** Example of mapping a TSN flow into a 5G QoS flow.

5G QoS flow, but at the cost of reduced granularity in QoS enforcement within the 5G system.

**Transparency of our VxLAN Solution to 5G Mobility Procedures**: In the context of a 5G-TSN-based industrial network, our VxLAN-based solution remains fully transparent to 5G handover procedures. On the UE side, as the VTEP is integrated directly into the UE, there is no need to create or delete any VxLAN when the UE moves from a source base station to a target base station, as the VTEP configuration remains unaffected by the base station change.

Focusing on the 5G core, only two UPFs are required within a single-site factory: one connected to the enterprise edge room and the other to an external network (e.g., the Internet), as shown in Fig. 2. Each of these UPFs is equipped with an integrated VTEP. An UE with mobility requirements could have two simultaneous PDU sessions: one directed to the edge room and another to the external network. Since there are only two UPFs, the PDU session towards the edge room will always be handled by the UPF connected to the edge room, and the PDU session towards the external network will always be handled by the UPF connected to the external network. As a result, during a handover procedure, there is no need to switch UPFs and thus, creating or deleting any VxLANs, as the respective PDU sessions maintain their dedicated endpoints in the same UPFs.

The configuration of the traffic forwarding through a VxLAN tunnel also remains unaffected by mobility. The forwarding table on the VTEPs (as shown in Fig. 2), which contains the destination IP addresses of the target UEs, remains unchanged regardless of mobility. The only change that occurs is the base station providing coverage to a UE with mobility requirements. When this UE moves to a new base station, the GPRS Tunneling Protocol (GTP) tunnel (responsible for carrying user data between the base station and the UPF) [24] must update its endpoint, shifting from the source base station to the target base station. However, it is important to note that the VxLAN tunnel is encapsulated within the GTP tunnel. The GTP tunnel is managed by the 5G mobility mechanisms defined by 3GPP [25], and any changes to the GTP tunnel during handovers do not impact the VxLAN tunnel, which remains transparent to these operations.

### B. EXAMPLE OF FRAME FORWARDING AND QOS TREATMENT

To illustrate the adopted solution, consider the example in Fig. 3, where an Ethernet unicast frame is transmitted from a centralized PLC to an actuator in production line #1. While this example illustrates a downlink unicast transmission, the following principles apply to uplink and broadcast scenarios.

The process begins with the Ethernet frame being transmitted through the TSN backbone (step 1). This transmission, within VLAN 100, undergoes QoS treatment at each TSN bridge based on PCP 5, e.g., frame scheduling using 802.1Qbv. At VTEP #X (step 2), the Ethernet frame is encapsulated into a VxLAN frame. The VNI 1005 is set using the tuple {VLAN ID, PCP}, with the first three digits for VLAN ID, and the last digit for PCP. The frame is further enclosed within an outer IP header, where the source IP address (192.168.1.100) is VTEP #X's address and the destination IP address (192.168.1.1) is VTEP #1's address, connected to production line #1. Additionally, the VTEP sets the DSCP value to 40, equivalent to PCP 5. Next, the UPF uses the PDR algorithm to filter the IP packet and map it into the appropriate QoS flow within the PDU session (step 3). The PDR aligns QFI #1 (5QI #86) with DSCP 40. Once assigned, the IP packet traverses the 5G Core (5GC) until it reaches the corresponding base station. According to the

5QI, the IP packet is associated with a specific DRB to ensure proper QoS treatment across the NR protocol stack. At VTEP #1 (step 4), the IP packet is decapsulated to restore the original Ethernet frame. Using the one-to-one mapping between VNI and {VLAN ID, PCP}, the egress VTEP can reinsert the VLAN and PCP tags, i.e., VLAN 100 and PCP 5, onto the original Ethernet frame in case they were removed by the ingress VTEP. Finally, the frame is routed within production line #1 to the target actuator (step 5).

## IV. PROOF OF CONCEPT
### A. TESTBED DESCRIPTION

To validate our solution, we implemented a testbed consisting solely of a commercial IP-based 5G system, as illustrated in Fig. 4. It comprises seven devices. A general-purpose computer equipped with a PCIe SDR50 card, referred to as 5G Amarisoft (Equip. 1), equipped with an Intel(R) Xeon(R) Bronze 3206R CPU @1.90GHz and 32GB RAM, runs the Amarisoft software to provide both the 5GC and RAN capabilities for a standalone 5G network. The testbed also includes two UEs, each consisting of an Intel NUC BXNUC10I7FNH2 (Equip. 2) paired with a Quectel RM500Q-GL card in an RMU500EK evaluation board (Equip. 3). This board uses the RM500QGLABR11A06M4G firmware and the NUC is equipped with an 11th Gen Intel(R) Core(TM) i7-1165G7 @2.80GHz and 16GB RAM. Both the 5G Amarisoft and UE operate on Ubuntu 18.04.6 LTS. Since this system works in licensed bands, it is enclosed in an RF Shielded Test Enclosure, specifically the Labifix LBX500 model (Equip. 4). The last component is a SecureSync 2400 time synchronization server (Equip. 5), which distributes time using the Network Time Protocol (NTP) to ensure time synchronization across devices.

### B. EXPERIMENTAL SETUP

We consider four traffic flows categorized by priority levels: High Priority (HP) with PCP 7, High-Medium Priority (HMP) with PCP 5, Low-Medium Priority (LMP) with PCP 2 and Low Priority (LP) with PCP 0. The HP flow considers a Precision Time Protocol (PTP) message transmitted every 125 ms. The HMP flow includes five cyclic Real Time Class 1 (RTC1) messages using the PROFINET protocol to set up five actuators, sent every 25 ms. The LMP flow carries augmented reality data via User Datagram Protocol (UDP) at 1 Mbps, while the LP flow involves UDP-based File Transfer Protocol (UFTP) data transmission at 9 Mbps. All flows are tagged with VLAN 100.

As depicted in Fig. 4, Ethernet frames for these flows are originated from the enterprise edge room. The HMP and LP flows are targeted at UE 1, while LMP flow is intended for UE 2. Additionally, the HP flow is directed to both UEs. We use the packETH tool[2] to insert these frames into the virtual network interface *veth1*. These frames are then redirected to the interface *veth0*, where the Linux `tc` tool is used for redirecting Ethernet frames to the corresponding VxLAN based on the tuple {VLAN ID, PCP}. Then, these frames are encapsulated to the corresponding VxLAN and tagged with an outer IP header with destination address corresponding to the considered UE. Next, these IP packets are inserted into the UPF network interface, i.e., *tun2*, and the Linux `iptables` command is used for setting a DSCP value based on the corresponding VxLAN. Later, the UPF maps each IP packet to a specific QoS flow based on the DSCP value and the base station maps such flows to specific DRBs. Once the IP packets arrive to the VxLAN interfaces at the UE side, they are decapsulated to recover the original Ethernet frames and they are redirected to the virtual network interface *veth0*, which emulates the endpoint of the production line. Note the VTEPs were implemented using the Linux `ip link` command. Contrary to the RFC 7348 standard, which does not require VTEPs to preserve VLAN and PCP tags during VxLAN encapsulation, `ip link` command retains the VLAN and PCP tags in the ingress VTEP (i.e., UPF). As a result, these tags are not reinserted during the decapsulation process in the egress VTEP (i.e., UE).

To foster research on the topic and favor reproducibility, we make configuration files and wireshark traces public in [26].

### C. PERFORMANCE RESULTS

To validate the proposed solution, we have used wireshark to capture traces from the described traffic flows. Specifically, we have captured traces from point A and D in Fig. 4 to demonstrate how the generated Ethernet frames are encapsulated to the appropriate VxLAN and they are mapped to the corresponding QoS flows. Observing the traces depicted in Fig. 5, we evidence the packets captured from point A are generated with specific tags for the VLAN and the PCP. In the traces captured from point D, we see each tuple {VLAN ID, PCP} is mapped to the corresponding VxLAN. Furthermore, the encapulated Ethernet frames are marked with the corresponding DSCP value at the IP header. Additionally, we observe each encapsulated Ethernet frame is mapped to a specific QFI. Note the QFI #1 corresponds to the LP flow, the QFI #2 corresponds to the LMP flow, the QFI #3 corresponds to the HMP flow, and the QFI #4 corresponds to the HP flow. Furthermore, we utilized the Amarisoft WebGUI to inspect how each QoS flow is mapped to a specific DRB at the base station (point E), ensuring that each flow receives a distinct priority for radio resource allocation. Additionally, we have captured traces from points G and H to demonstrate the frames reach the proper destination. Specifically, we have observed the multicast PTP frame (i.e., from HP flow) reaches both UEs whereas the remaining unicast frames arrive to the proper UE.

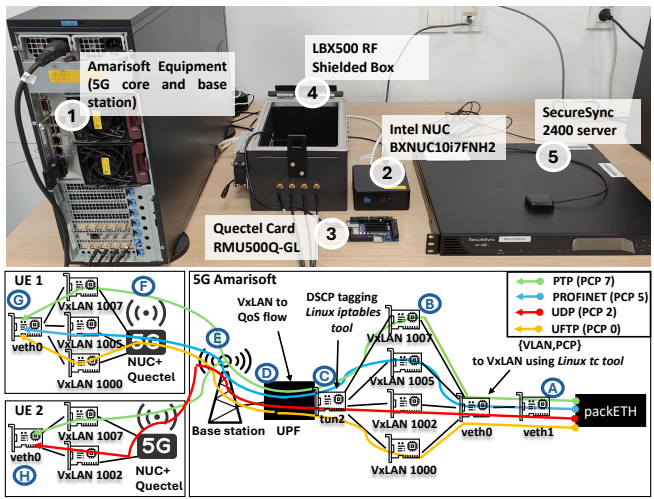

**FIGURE 4.** Proof of concept equipment and evaluated network scenario. To avoid redundancy, only one Intel NUC and one Quectel card, representing a single UE, are depicted in the top image.

---

[2]Packet Generator Tool (PackETH): https://github.com/jemcek/packETH

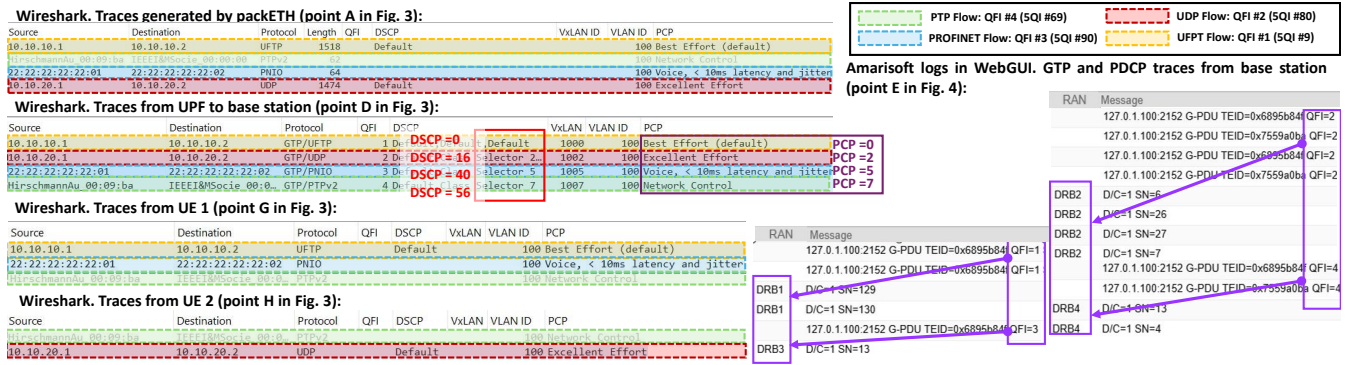

**FIGURE 5.** Wireshark traces and Amarisoft logs from the network scenario described in Fig. 4.

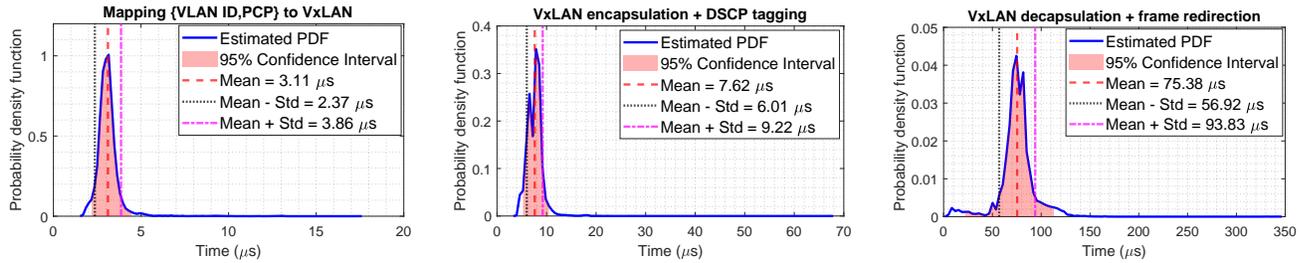

**FIGURE 6.** PDFs of the execution delay of various VTEP tasks within the 5G system. Task 1: Mapping {VLAN ID, PCP} to VxLAN (left plot). Task 2: VxLAN encapsulation + DSCP tagging (middle plot). Task 3: VxLAN decapsulation + frame redirection (right plot).

We have also conducted measurements to assess the delay caused by various VTEP tasks within the 5G system. By measuring these delays, we aim to evaluate their impact on packet transmission delay in the 5G system.

- Task 1 involves mapping the tuple {VLAN ID, PCP} to a VxLAN, essentially redirecting Ethernet frames to the corresponding VxLAN interface (from point A to B in Fig. 4).
- Task 2 encompasses VxLAN encapsulation along with DSCP tagging (from point B to C).
- Task 3 entails VxLAN decapsulation and the redirection of recovered Ethernet frames (from point F to G).

Our analysis considers the transmission of approximately 1 million of Ethernet frames based on the described traffic flows. We utilized PCAP files generated by Wireshark to measure the delays of the previously described tasks, specifically using timestamps with nanosecond precision and the sequence numbers of the Ethernet frames [27]. It yields the following results:

The Probability Density Functions (PDFs) of the execution delay for Tasks 1, 2, and 3, along with the 95% confidence interval, are shown in Fig. 6, also highlighting the mean values and standard deviations. Task 1 introduces an average delay of 3.1130 $\mu s$ with a standard deviation of 0.7476 $\mu s$ and a 95% confidence interval of [2.1120 4.4990] $\mu s$. Task 2 exhibits an average delay of 7.6193 $\mu s$ with a standard deviation of 1.6049 $\mu s$ and a 95% confidence interval of [4.8640 10.6190] $\mu s$. Finally, Task 3 shows an average delay of 75.3753 $\mu s$ with a standard deviation of 18.4595 $\mu s$ and a 95% confidence interval of [20.4410 115.0144] $\mu s$.

Note the faster processing in Tasks 1 and 2 is attributed to the superior CPU in the PC running Amarisoft, compared to the NUC which executes Task 3. These findings indicate that the total delay introduced by these tasks is approximately 100 $\mu s$ on average, which is negligible when compared to the typical packet transmission delay of a 5G system, which typically ranges from a few milliseconds to tens of milliseconds.

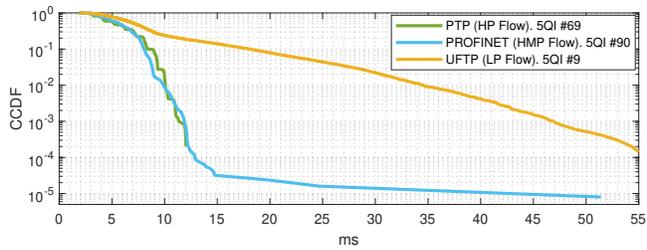

**FIGURE 7.** CCDF of the transmission delay of a packet from point A to point G for traffic flows directed to UE 1 (see Fig. 4).

Fig. 7 presents the Complementary Cumulative Distribution Function (CCDF) of the transmission delay of a packet for HP, HMP, and LP flows. Specifically, it shows that the packet delay budget is met for HP/LP flows, while the HMP flow stays within the packet delay budget in 99.998% of cases. These results also confirm that the traffic prioritization, based on the PCP values used in TSN networks, is preserved in the 5G system. This means these results demonstrate our solution effectively harmonizes the QoS mechanisms of the 5G system with those of a TSN network.

Note the focus of our study is not to optimize uRLLC mechanisms in the 5G system to minimize latency, as achieving such optimization is a complex challenge that requires an extensive analysis of several parameters, including flexible sub-carrier spacing, mini-slot scheduling, semi-static scheduling, PUSCH enhancements, sub-slot-based HARQ-ACK feedback, logical channel prioritization restrictions, and intra-UE prioritization [28]. A detailed study on how to optimally configure these mechanisms is beyond the scope of this work.

## V. CONCLUSIONS AND FUTURE WORKS

This paper explores the integration of an IP-based 5G system with a TSN industrial network, focusing on Ethernet frame forwarding among interconnected TSN islands via 5G. We also address QoS harmonization between 5G and TSN. Due to the lack of commercial 5G UEs supporting Ethernet-based sessions, we adopt a VxLAN-based solution to encapsulate Ethernet frames within IP-based 5G systems, preserving Ethernet header information. VLAN IDs and PCP values are mapped to VNIs for efficient frame forwarding, while PCP values are translated to 5QIs to ensure consistent QoS across TSN and 5G domains. The latter involves translating PCP to DSCP in the IP header during VxLAN encapsulation, allowing packet filters to assign packets to appropriate 5G QoS flows. Our solution is validated with a prototype, showing effective traffic classification and forwarding. The VxLAN mechanisms, including VxLAN header insertion and PCP-to-DSCP mapping, introduce an average latency of approximately 100 $\mu s$, which is negligible compared to typical 5G packet transmission delays ranging from a few to tens of milliseconds. Furthermore, our results demonstrate the proposed solution preserves QoS treatment between the 5G system and TSN, ensuring the priority of 5G QoS flows remains aligned with the PCP priorities of industrial traffic flows.

Future work will expand the testbed to explore TSN mechanisms like Time-Aware Shaping (TAS) in a full 5G-TSN network, focusing on performance, scalability and synchronization message transmission.

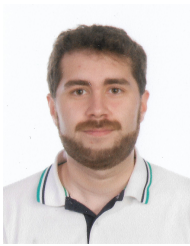

**OSCAR ADAMUZ-HINOJOSA** received the B.Sc., M.Sc., and Ph.D. degrees in telecommunications engineering from the University of Granada, Granada, Spain, in 2015, 2017, and 2022, respectively. He was granted a Ph.D. fellowship by the Education Spanish Ministry in September 2018. He is currently an Interim Assistant Professor with the Department of Signal Theory, Telematics, and Communication, University of Granada. He has also been a Visiting Researcher at NEC Laboratories Europe on several occasions. His research interests include network slicing, 6G radio access networks (RAN), and deterministic networks, with a focus on mathematical modeling.

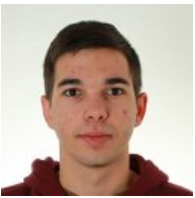

**FELIX DELGADO-FERRO** received the B.Sc. and M.Sc. degrees in telecommunication engineering from the University of Granada, Spain, in 2020 and 2022. Currently, he is developing his doctoral thesis in telecommunication engineering with the Wireless and Multimedia Networking Lab (WiMuNet) Research Group in the Department of Signal Theory, Telematics, and Communications, University of Granada. In 2023, he was a visiting doctoral student at the Simula Research Laboratory (SINLab), University of Oslo, Norway. His current research interests include multipath protocols, heterogeneous networks, and optimization techniques for IoT, 5G, and 6G networks.

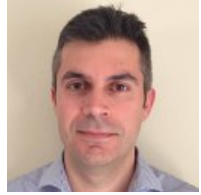

**JORGE NAVARRO-ORTIZ** received the M.Sc.E.E. degree from the University of Malaga, Spain, in 2001. Then, he worked at Nokia Networks, Optimi/Ericsson, and Siemens. He started working as an Assistant Professor at the University of Granada, in 2006, where he got the Ph.D degree in 2010. He is currently an Associate Professor with the Department of Signal Theory, Telematics and Communications, University of Granada. His research interests include wireless technologies for the IoT, such as LoRaWAN and 5G.

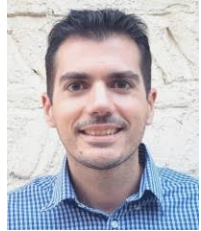

**PABLO MUÑOZ** received the M.Sc. and Ph.D. degrees in telecommunication engineering from the University of Málaga, Málaga, Spain, in 2008 and 2013, respectively. He is currently an Associate Professor with the Department of Signal Theory, Telematics, and Communications, University of Granada, Granada, Spain. He has published more than 50 articles in peer-reviewed journals and conferences. He is the coauthor of four international patents. His research interests include radio access network planning and management, the application of artificial intelligence tools in radio resource management, and the virtualization of wireless networks.

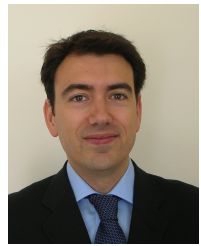

**PABLO AMEIGEIRAS** received the M.Sc.E.E. degree from the University of Málaga, Málaga, Spain, in 1999. He carried out his Master thesis at the Chair of Communication Networks, Aachen University, Aachen, Germany. In 2000, he joined Aalborg University, Aalborg, Denmark, where he carried out his Ph.D. thesis. In 2006, he joined the University of Granada, Granada, Spain, where he has been leading several projects in the field of 4G and 5G systems. His research interests include 5G, 6G, the IoT, and deterministic networks.